\documentclass[twocolumn]{aastex61}
\pdfoutput=1 
\usepackage{amsmath,amstext}
\usepackage[T1]{fontenc}
\usepackage{apjfonts} 
\usepackage[figure,figure*]{hypcap}
\usepackage{multirow}
\usepackage{booktabs}

\newcommand{\fermi}{\textit{Fermi}}

\usepackage{graphicx}

\newcommand{\commenty}[1]{\textcolor{black}{#1}}

\newcommand{\package}[1]{\textsc{#1}}

\shorttitle{GMRT view of GRB 170817A/GW 170817}
\shortauthors{Resmi et al}

\begin{document}
\title{Low frequency view of GW 170817/GRB 170817A with the Giant Metrewave Radio Telescope}

\author{L. Resmi}
\affiliation{Indian Institute of Space Science \& Technology, Trivandrum 695547, India}

\author[0000-0001-6797-1889]{S. Schulze}
\affiliation{Department of Particle Physics and Astrophysics, Weizmann Institute of Science, Rehovot 761000, Israel}

\author{C.~H. Ishwara-Chandra}
\affiliation{National Center for Radio Astrophysics, Pune 411007, India}
\affiliation{Inter-University Institute of Data Intensive Astronomy, and Dept. of Astronomy, Univ. of Cape Town, Cape Town 7700, South Africa}

\author[0000-0003-1637-267X]{K. Misra}
\affiliation{Aryabhatta Research Institute of observational sciencES (ARIES), Manora Peak, Nainital 263 001, India}

\author{J. Buchner}
\affiliation{Pontificia Universidad Cat{\'{o}}lica de Chile, Instituto de Astrof{\'{\i}}sica, Casilla 306, Santiago 22, Chile} 

\author{M. De Pasquale}
\affiliation{Department of Astronomy and Space Sciences, Istanbul University, 34119 Beyaz\i t, Istanbul, Turkey} 

\author[0000-0002-7158-5099]{R.~S\'{a}nchez-Ram\'{\i}rez}
\affiliation{INAF, Istituto di Astrofisica e Planetologia Spaziali, Via Fosso del Cavaliere 100, I-00133 Roma, Italy}
\affiliation{Instituto de Astrof\'{\i}sica de Andaluc\'{\i}a (IAA-CSIC), Glorieta de la Astronom\'{\i}a s/n, E-18008, Granada, Spain.}

\author{S. Klose}
\affiliation{Th\"uringer Landessternwarte, Sternwarte 5, 07778 Tautenburg, Germany}

\author[0000-0001-8332-0023]{S. Kim}
\affiliation{Instituto de Astrof{\'{\i}}sica and Centro de Astroingenier{\'{\i}}a, Facultad de F{\'{i}}sica, Pontificia Universidad Cat{\'{o}}lica de Chile, Casilla 306, Santiago 22, Chile}
\affiliation{Max-Planck-Institut f\"{u}r Astronomie K\"{o}nigstuhl 17 D-69117 Heidelberg, Germany}

\author[0000-0003-3274-6336]{N. R. Tanvir}
\affiliation{Department of Physics and Astronomy, University of Leicester, University Road, Leicester, LE1 7RH, United Kingdom}

\author{P. T. O'Brien}
\affiliation{Department of Physics and Astronomy, University of Leicester, University Road, Leicester, LE1 7RH, United Kingdom}

\correspondingauthor{L. Resmi}
\email{l.resmi@iist.ac.in}


\begin{abstract}
The short gamma-ray burst (GRB) 170817A was the first GRB associated with a gravitational-wave event. Due to the exceptionally low luminosity of the prompt $\gamma$-ray and the afterglow emission, the origin of both radiation components is highly debated. 
The most discussed models for the burst and the afterglow include a regular GRB jet seen off-axis and the emission from the cocoon encompassing a "choked" jet. Here, we report  low radio-frequency observations at 610 and 1390~MHz obtained with the Giant Metrewave Radio Telescope (GMRT). Our observations span a range of $\sim7$ to $\sim152$ days after the burst. The afterglow started to emerge at these low frequencies about 60~days after the burst. The $1390$~MHz light curve barely evolved between 60 and 150 days, but its evolution is also marginally consistent with a $F_\nu\propto t^{0.8}$ rise seen in higher frequencies. We model the radio data and archival X-ray, optical and high-frequency radio data with models of top-hat and Gaussian structured GRB jets. We performed a Markov Chain Monte Carlo analysis of the structured-jet parameter space. Though highly degenerate, useful bounds on the posterior probability distributions can be obtained. Our bounds of the viewing angle are consistent with that inferred from the gravitational wave signal. We estimate the energy budget in prompt emission to be an order of magnitude lower than that in the afterglow blast-wave. 
\end{abstract}

\section{Introduction}

The Laser Interferometer Gravitational-Wave Observatory (LIGO) and VIRGO gravitational wave (GW) detectors detected on 2017 August 17 for the first time the emission from two inspiraling neutron stars \citep[GW170817;][]{LIGO_VIRGO_1}. The 3-dimensional localization inferred from the GW signal enabled a global network of observers to detect for the very first time electromagnetic radiation emitted during and after the neutron star inspiral.

About 1.7~s after the beginning of the neutron star inspiral, the $\gamma$-ray satellite \fermi\ detected a short gamma-ray burst GRB\,170817A that also coincided spatially with GW170817 \citep{Goldstein2017b}. 
As soon as the field became visible from the ground, optical and near IR observations detected a new source in the credible region of GW170817, dubbed AT2017gfo \citep[e.g.,][for a review see also \citealt{MMA2017a}]{Arcavi2017a, Coulter2017a, Lipunov2017a, SoaresSantos2017, Tanvir2017b}. Radio, sub-mm and X-ray observations revealed no counterpart during the first week after GW170817 \citep{Alexander2017a, Evans2017a, Hallinan2017b, Kim2017, Margutti2017b}. Only 9 days after GW170817, a new source at the position of AT2017gfo emerged at X-ray frequencies \citep{Troja2017b} and a week later also at radio frequencies \citep{Hallinan2017b}.

Modeling the multiband data revealed two distinct phenomena powering the long-lasting emission from radio to X-ray frequencies. The UV-to-NIR emission up to $\sim30$~days originated from the radioactive decay of lanthanides \citep[e.g.,][for a critical reflection see also \citealt{Waxman2017a}]{Pian2017a, Tanvir2017b, Evans2017a}. The emission at longer and shorter wavelengths is of non-thermal origin. The brightness of this component increased since the discovery \citep[$F_\nu\propto t^{0.8}$;][]{Haggard2017b, Hallinan2017b, Troja2017b, Mooley2017a, Margutti2017b}, while the shape of the spectral energy distribution remained constant with time \citep[$F_\nu\propto\nu^{-0.6}$;][]{Mooley2017a}.
About 110 days after GW170817, long after the kilonova faded, this source also started to emerge at optical wavelengths \citep{Lyman2018a, Margutti2018a}.

Since GW170817 was accompanied by a short-duration GRB, the non-thermal component might naturally  be connected with the GRB afterglow but seen off-axis \citep{Granot2017a, Margutti2017b, Kim2017, Troja2017b}. However, \citet{Kasliwal2017a}, \citet{Mooley2017a} and \citet{Nakar2018a} argued that the emission could be produced by the low-luminosity sub-relativistic cocoon. Recently, \citet{Hotokezaka2018a} proposed that the observed non-thermal emission could also be produced by the interaction of the fast tail of the neutron star ejecta with the circumstellar material.


In this paper, we present our continuing low-frequency observations of the radio transient using the Giant Metrewave Radio Telescope (GMRT), located in Pune, India ($\S$\ref{sec:gmrt_observations}), covering the time interval from 7 to 152~days after GW170817. The transient started to emerge at 1390 MHz frequencies $\sim67$~days after GW170817 and $\sim40$~days later also at 610~MHz. We augment our data set with archival X-ray and radio data to model the evolution in the framework of a structured GRB jet. To characterize the highly degenerate multi-dimensional parameter space of the model, we applied the Markov Chain Monte Carlo (MCMC) technique. 
All uncertainties in this paper are quoted at $1\sigma$~confidence. We assume the distance to GW170817 to be 42.5 Mpc \citep{Hjorth2017a}.

\section{Observations and data reduction}\label{sec:gmrt_observations}

\subsection{GMRT observations}

We began monitoring the afterglow of GRB 170817A with the Giant Metrewave Radio telescope \citep[GMRT;][]{Swarup1991a} around a week after the burst \citep{Resmi2017a}. The observations up to 30 days were carried out at the L-band  with the 32-MHz legacy correlator at 1390 MHz. These early observations yielded only upper limits \citep{Kim2017} (hereafter Paper-I). We continued our observations through a series of Director's Discretionary Time (DDT) proposals (PI Kuntal Misra), using the upgraded GMRT. On 23 October 2017, $67$ days past the burst at $1390$~MHz, we secured the first detection at 1390 MHz.
Each of the observations took about $\sim4$~hours, including overheads for calibration and slewing. A log of our observations is shown in Table \ref{gmrttable}.


\subsection{Data analysis}


We processed the wideband data with the \package{Common Astronomy Software Applications} (\package{CASA}\footnote{\href{https://casa.nrao.edu}{https://casa.nrao.edu}}) package  \citep{McMullin2007a} and the legacy system  data with \package{NRAO Astronomical Image Processing Software} (\package{AIPS}\footnote{\href{http://www.aips.nrao.edu}{http://www.aips.nrao.edu}}) package \citep{Wells1985a}. The data were flagged and calibrated using standard procedures.  The primary calibrators 3C286 or 3C147 were used as flux and bandpass calibrators and  J1248$-$199 was used for phase calibration. After flux, gain and bandpass calibration, the channel averaging was done to the extent to minimize the effect of bandwidth smearing and target was split.
 On the target, we performed a few rounds of  phase-only self-calibration and afterward a few rounds of amplitude and phase self-calibration.

The beam at 610~MHz has a radius of $\gtrsim5''$ and is comparable to the distance from AT2017gfo to the  nucleus of its host galaxy. 

The $610$ and $1390$ MHz flux was measured with a two-component Gaussian fit (for peak and underlying baseline) on the final images using \package{JMFIT} in \package{AIPS}, centered at the afterglow and the galaxy nucleus. {\commenty{The final error quoted is the quadrature sum of the (i) map RMS, (ii) JMFIT error, and (iii) the flux scale error measured as the uncertainty in the calibrator flux.}} Transient and host flux measurements are summarized in Table \ref{gmrttable}. 



\subsection{GMRT Light curve}

\begin{figure}
\begin{center}
\includegraphics[width=1\columnwidth]{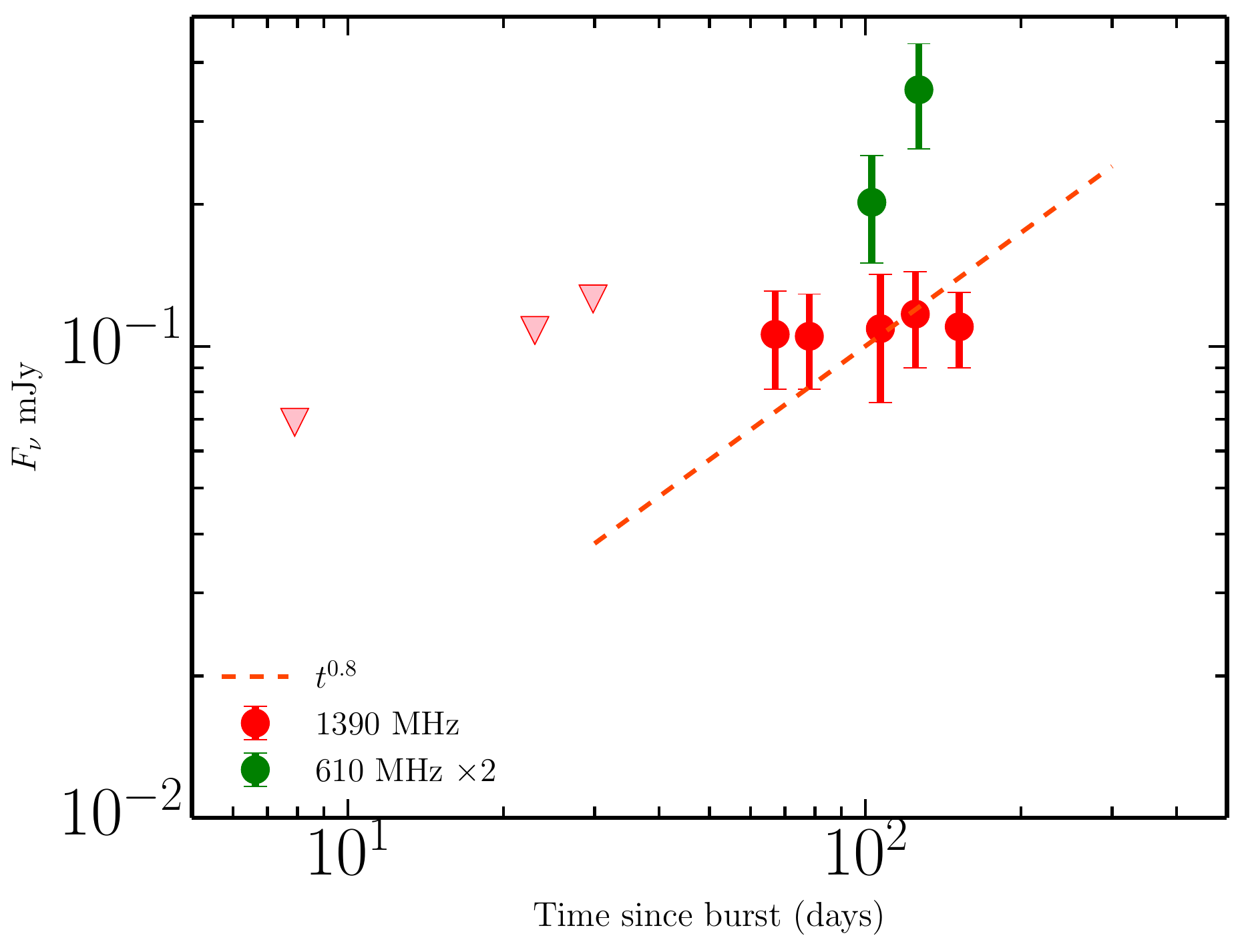}
\caption{The GMRT observations of the afterglow of GRB 170817A. The observations we report in the paper are given by solid symbols. The $610$~MHz points are shifted up by a factor or two. Upper limits are displayed as downward-pointing triangles and were previously reported in Paper-I. The dashed line represents a $t^{0.8}$ rise seen in higher-frequency radio data, over-plotted with $1390$~MHz data.}
\label{figgmrt}
\end{center}
\end{figure}

Taken in isolation, the GMRT $1390$~MHz lightcurve is consistent with a plateau phase (Fig. \ref{figgmrt}). However, it is also marginally consistent with the $t^{0.8}$ rise previously derived from X-ray and high radio frequency observations \citep{Mooley2017c, Margutti2018a}. Our last L-band lightcurve extends $152$~days after GW\,170817, while the high radio frequency observations extend to $115$~days. The flatness could indeed be a slow turn-over of the afterglow lightcurve, which was not apparent in the high radio frequency observations reported so far {\footnote{\commenty{This corresponds to the time of the submission of the paper. However, data released after the submission of the paper confirm the turn-over.}}}. This conclusion is also consistent with the late time X-ray observations reported by \citet{Troja2018b}, where the \textit{Chandra} flux at $158$ days is consistent with that at $110$ days since the burst. 

Variabilities could also provide an explanation for the plateau phase (see Fig. \ref{figgmrt}). In X-ray frequencies also the flux is found to be variable \citep{Troja2018b}.

The $610$~MHz lightcurve is consistent with both a plateau and a $t^{0.8}$ rise. However, it spans only for a duration of $\sim 25$ days. 


\subsection{Archival data}

We augment our data set with X-ray measurements reported in \citet{Margutti2018a}, \citet{DAvanzo2018a} and \citet{Troja2018a}, optical observations in reported \citet{Lyman2018a}, and radio observations reported  in \citet{Mooley2017c}, \citet{Margutti2018a} and \citet{Troja2018a}.


\begin{table}
\caption{Log of GMRT observations of GRB 170817A}
\begin{small}
\centering
\begin{tabular}{cccccc}
\hline
Date 	& $t-t_0$ 	& $F_\nu$ (AG) & $F_\nu$ (host) 	& RMS &beam size\\ 
		& (days) 	& ($\mu$Jy) & ($\mu$Jy)& ($\mu$Jy)& (arcsec) 	\\
\hline
\multicolumn{6}{c}{\textbf{600 MHz}}\\
\hline
28-11-2017 &  102.5 	& {\commenty{$101 \pm 26$}}	&{\commenty{$1003 \pm 23$}}	& {\commenty{$16$}}& $5.61 \times 5.26$ \\
22-12-2017 &  127.0 	&{\commenty{$175 \pm 44$}}	&{\commenty{$973 \pm 33$}}	& {\commenty{$20$}}& $5.98 \times 3.59$ \\
\hline
\multicolumn{6}{c}{\textbf{1390 MHz}}\\
\hline
23-10-2017 &  66.6 	& {\commenty{$106 \pm 25$}} &{\commenty{$721 \pm 60$}}& {\commenty{$19$}} & $4.41 \times 3.05 $\\ 
03-11-2017 &  77.6 	& {\commenty{$ 105 \pm 24$}} &{\commenty{$769 \pm 62$}}& {\commenty{$17$}}&$2.61 \times 2.19 $\\ 
02-12-2017 &  106.6 	& {\commenty{$109 \pm 33$}} & {\commenty{$849 \pm 83$}}& {\commenty{$30$}}&$3.82 \times 2.45 $\\
20-12-2017 &  124.5 	& {\commenty{$117 \pm 27$}} & {\commenty{$851 \pm 47$}}& {\commenty{$19$}} &$2.87 \times 2.16$ 				\\
16-01-2018 &  151.5 	& {\commenty{$110 \pm 20$}} & {\commenty{$755 \pm 38$}}& {\commenty{$14$}} &$2.83 \times 2.15$ \\
\hline
\end{tabular}
\label{gmrttable}
\end{small}
\end{table}


\section{Modelling}\label{sec:modelling}
\subsection{Uniform top-hat jet model}
In Paper-I, we used the results from the high-resolution two-dimensional relativistic hydrodynamical code \package{BOXFIT} version 2 \citep{vanEerten2012a} to interpret the multi-band afterglow under the ambit of the uniform top-hat jet model. Along with the data, in that paper we presented two out of the several plausible solutions: (i) a narrow jet of half-opening angle $\sim 5^{\circ}$ misaligned at $\sim 17^{\circ}$ from the observer, and (ii) a wide jet of opening angle $\sim 20^{\circ}$ with the jet axis $41^{\circ}$ away from the observer line of sight.


However, recent observations indicated that the top-hat jet model is insufficient to explain the behavior of the afterglow \citep{Mooley2017c, Margutti2018a}. We found that the lateral expansion of the jet led to a steep decay after the peak in top-hat jet models that could fit the early data of the afterglow. We found that an $F_\nu\propto t^{0.8}$ rise lasting for about a decade in time is possible with a uniform top-hat jet. However, the parameters that lead to a $F_\nu\propto t^{0.8}$ phase between $10 - 100$ days require unreasonably high values of energy and density of the ambient medium, which also lead to fluxes several orders of magnitude larger than observed. A low fraction of accelerated electrons could reduce the flux, but that results in high values of $\gamma_m$ leading to the disagreement with the observed spectrum of the afterglow.


%

\subsection{Structured-jet model}\label{sec:model}
In order to explain the evolution of the afterglow, several groups have invoked   either a radial \citep{Mooley2017c} or a lateral \citep{Lazzati2017a, Margutti2018a, Lyman2018a} structure in the energy and velocity profile of the outflow. In the radially structured cocoon model, the relativistic jet is "choked" and the burst and the afterglow originate from a sub-relativistic cocoon \citep{Mooley2017c, Nakar2018a}.  


Since evidence for relativistic jets are seen in gamma ray bursts \citep{Frail:1997qf,Taylor:2004wd}, we investigate the parameter space of a structured relativistic jet to explain the afterglow observations. The Gaussian structured jet we consider is similar to previously discussed by \citet{Lazzati2017a}, \citet{Margutti2017a}, \citet{DAvanzo2018a}, \citet{Lyman2018a}, and \citep{Troja2018c} where the kinetic energy per solid angle ($\cal{E}$) has a polar structure given by
\begin{equation}
	{\cal{E}}(\theta) = E_c \exp{\left( -\frac{\theta^2}{\theta_{c}^2} \right)},
\label{eq1}
\end{equation}
where $\theta_c$ is the {\textit{structure parameter}} deciding the sharpness of the angular profile. A jet with a large $\theta_c$ is similar to a uniform jet.  To have the same deceleration radius ($r_0$) across the polar direction, we let the initial bulk Lorentz factor to follow,
\begin{equation}
\Gamma_0 \beta_0 = \Gamma_c \beta_c \exp{\left(- \frac{\theta^2}{2 \theta_{c}^2} \right)},
\end{equation}
where $\beta$ is the bulk velocity of the jet normalized by the speed of light and $\Gamma$ is the bulk Lorentz-factor of the GRB jet. A jet where the kinetic energy is proportional to $\exp\left(-\theta^2\right)$ and the Lorentz factor is proportional to $\exp\left(-\theta^2/2\right)$ is possible if the ejected mass also follows an angular profile of $\exp\left(-\theta^2/2\right)$, which is a 
reasonable assumption to make. 

Due to the angular structure, the initial jet Lorentz factor decreases toward high latitudes and becomes sub-relativistic toward high latitudes. We assume that the initial jet Lorentz factor follows
\begin{equation}
\Gamma (\theta) \beta (\theta) = \Gamma_0 (\theta) \beta_0 (\theta) \left( \frac{r}{r_0} \right)^{-3/2}
\end{equation}
where $r \gg r_0$ can be considered equal to the distance from the center of the explosion. This velocity profile reduces to the Blandford-McKee self-similar solution in the ultra-relativistic limit \citep{Blandford1976a} and the Sedov-von Neumann-Taylor solution in the non-relativistic limit \citep{Taylor1950a}. 
To simplify the calculation, we assume a rigid jet that does not expand laterally.

To calculate the flux observed by an observer at an angle $\theta_v$ from the jet axis, we follow the same formalism developed by \citet{LK17}. We divide the jet into $N$  polar rings of width $\delta \theta = \theta_j/N$, which are further divided into $M$ azimuthal elements, each with a width of $\delta \phi = 2 \pi/M$. The direction to the observer's line of sight is taken as the zero of the $\phi$ coordinate. An element ${i,k}$ with its central axis at $(\theta_i, \phi_k)$ from the jet axis, is at an inclination $\alpha_{i,k} = \cos{\theta_i} \cos{\theta_v} + \sin{\theta_i} \sin{\theta_v} \cos{\phi_k}$ from the observer. As done in \citet{LK17}, we sum the contribution of each of these elements to obtain the total flux observed by the off-axis observer at $t_{\rm obs}$ at a frequency $\nu_{\rm obs}$. For this we interpolate the equation $t_{\rm obs} = \frac{r}{\beta(r) c} \left[ 1 - \beta(r) \cos{\alpha_{i,k}}\right] $ and find the distance $r$ corresponding to the $t_{\rm obs}$ for each jet element ${i,k}$.  The off-axis flux from each element is estimated as $a^3 {f^{\rm on}}_{\nu/a,i,k} (r)$, where $a = \frac{1-\beta(r)}{1-\beta(r) \cos{\alpha_{i,k}}}$ and $f^{\rm on}$ is the flux observed by an observer located on the central axis of the element. Again, we follow \citet{LK17} to obtain ${f^{\rm on}}_{\nu,i,k}$ as $\frac{L_{\nu,i,k}}{4 \pi d_L^2} \frac{\Omega_{i,k}}{\Omega_{e,i,k}}$. Here $\Omega_{e,i,k} =\max\left[ \Omega_{i,k}, 2 \pi (1 - \cos{\frac{1}{\Gamma_{i,k}}}) \right]$ and $\Omega_{i,k} = \int_{\phi_{k-1}}^{\phi_k} d\phi \int_{\theta_{i-1}}^{\theta_i} d\theta \sin{\theta}$ is the solid angle subtended by an element at a point on its central axis.

The isotropic synchrotron flux $L_{\nu}/4 \pi d_L^2$ is estimated following \citet{Sari1998a}, with modifications for expressions of the downstream magnetic field, $B$, and minimum Lorentz factor, $\gamma_m$, of the shocked electron population, suitable for the sub-relativistic flow: $B = \sqrt{32 \pi m_p c^2 \epsilon_B n_0  \Gamma (\Gamma-1) }$, and $\gamma_m  = 1+ \frac{m_p}{m_e} \frac{p-2}{p-1}\epsilon_e (\Gamma -1)$. Here, $m_p$ and $m_e$ are the proton and electron masses, respectively, $c$ is the speed of light, $p$ is the power-law index of the non-thermal electron population, and $\epsilon_e$ and $\epsilon_B$ are the fractional energy content in the non-thermal electron population and magnetic field, respectively. This formalism assumes a geometrically and optically thin jet. A thin shell assumption is valid because the broad-band spectrum even at MHz frequencies shows no signs of self-absorption to date.

Our model has minor differences from other structured-jet models presented in the literature. The angular profile of ${\cal E}$ and $\Gamma \beta$ of \citet{DAvanzo2018a} is different from ours, but they use the same dynamical evolution for $\Gamma \beta$. Both \citet{DAvanzo2018a} and \citet{Margutti2018a} use a free index ($s_1$ and $\alpha$, respectively) to modify the angular structure. Since the afterglow parameter space is heavily degenerate with at least $7$ free parameters, we chose to fix the Gaussian profile. The angular profile we use is very similar to \citet{LK17} except we let ${\cal E} \propto \exp\left(-\frac{\theta^2}{\theta_c^2}\right)$ in order to have a $r_0$ independent of the jet latitude. Our calculation of the jet dynamics and the equal arrival times differ from \citet{LK17}. They used a $t^{-3/8}$ profile for $\Gamma$ and scaled the observed time by the Doppler factor ($a$) to obtain the equal arrival times. Since $\beta$ can be considerably lower than unity {\commenty{at high jet latitudes}}, we chose to do an interpolation to obtain the equal arrival time surfaces. 

\subsection{Model parameters and the shape of the lightcurve}

The structured jet can be specified with four parameters, $E_c, \Gamma_c, \theta_c$, and $\theta_j$. The first three parameters were mentioned in the previous section. The last parameter $\theta_j$ corresponds to the half-opening angle if there is a hard edge of the jet beyond which the energy sharply drops down. 

\begin{figure*}
\begin{center}
\includegraphics[width=1\textwidth]{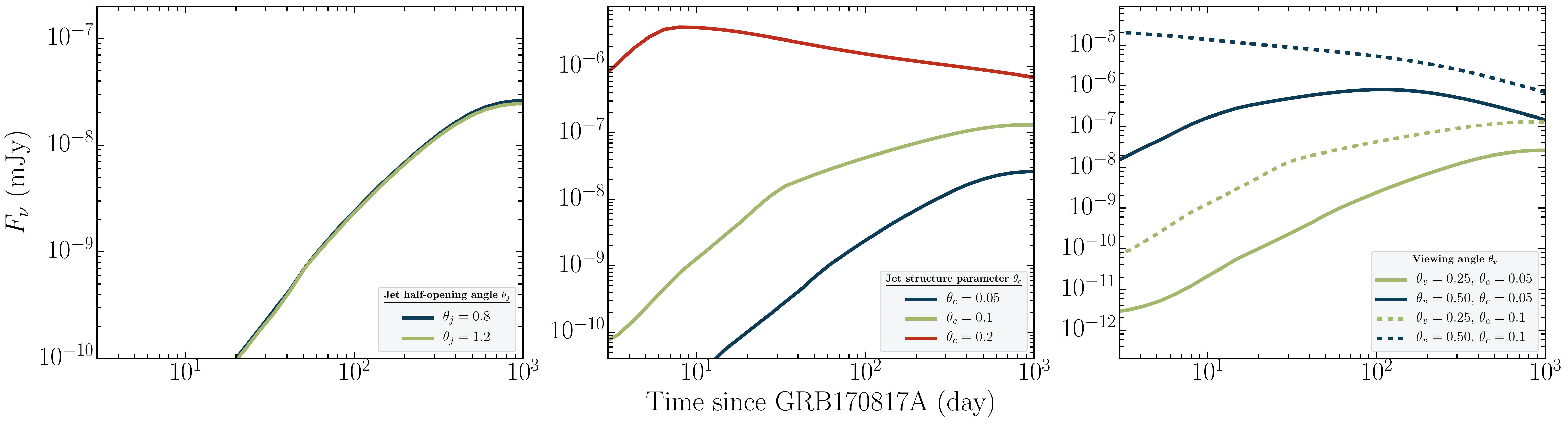}
\end{center}
\caption{The Sensitivity of lightcurves (3~keV) on the structured-jet parameters. \textbf{Left}: Variation in the jet half-opening angle $\theta_j$. An increase in the jet half-opening angle hardly changes the total flux. At large jet angles, the energy content in the Gaussian jet is negligible compared to the central part. For both lightcurves we assumed $\theta_c = 0.05$~rad and $\theta_v = 0.5$~rad.
\textbf{Middle}: Variation of jet structure  parameter $\theta_c$. The slope of the lightcurve before the peak is sensitive to $\theta_c$. The peak time is determined by $\theta_v$ and is the same for the blue and the green models. As $\theta_c$ gets larger, the lightcurve approaches to that of a uniform jet, and the peak shifts to earlier times. Since all curves have the same energy per solid angle (${\cal{E}}= 5\times 10^{51}$), an increase in $\theta_c$ leads to higher $E_{\rm tot}$, which is the reason for the higher flux. For these lightcurves, we assumed $\theta_j = 0.8$ rad and $\theta_v  = 0.5$ rad.
\textbf{Right}: Effect of viewing angle for two different $\theta_c$ values. A larger viewing angle leads to a later peak. As seen in the middle panel, the pre-peak slope is sensitive to $\theta_c$. The rising phase of the blue dashed curve is steep, but it is out of the x-axis range. 
}
\label{variation}
\end{figure*}

Similar to top-hat models, $\Gamma_c$ influences the deceleration time of the jet, and $E_c$ influences both the deceleration time as well as the overall level of flux. Hence we concentrate here on the jet profile $\theta_c$ and the jet half-opening angle $\theta_j$. In addition, the observer's viewing angle $\theta_v$ also modifies the lightcurve. Figure \ref{variation} shows a diverse assembly of lightcurves for different values of $\theta_j, \theta_c,$ and $\theta_v$.

The half-opening angle $\theta_j$ is the least sensitive of all. 
The jet structure parameter $\theta_c$ plays a crucial role in the rise time and the slope of the lightcurve. In addition, along with $\theta_v$, $\theta_c$ also influences the peak time. A small $\theta_c$ corresponds to a sharply varying profile, where the jet core is far more energetic than its edges, whereas a large $\theta_c$  broadly resembles a uniform jet.


\subsection{Parameter estimation}
\label{mcmc}



The parameter space of our afterglow model is 9-dimensional, $\Theta = E_c, \Gamma_c, \theta_c, \theta_j, \theta_v, n_0, \epsilon_B,\epsilon_e, p$. To reduce the number of free parameters, we make two additional assumptions. Firstly, the spectral energy distribution from X-ray to radio frequencies is well described by a simple power law. Following \citet{Margutti2018a}, we fix the value to $p=2.17$. Second, it is natural to let the profile go toward a polar angle of $\pi/2$. Moreover, we found that the lightcurves are not very sensitive to the value of $\theta_j$ (left panel of Fig. \ref{variation}). No bounds on $\theta_j$ could be obtained from our initial MCMC analysis either. Therefore, we fixed $\theta_j = 1.2$~rad in our final analysis. We chose a value less than $\pi/2$ to avoid numerical errors in the synchrotron flux calculation that result from extremely low values of jet velocity $\beta$. We thus reduced the number of free fitting parameters to 7.


\begin{table}
\caption{Prior and posterior distributions of the structured-jet afterglow model}
\begin{center}
\begin{tabular}{ccc}
Parameter 				& Prior range	& Posterior\\ 
\hline
$\log (E_c/{\rm erg})$ 		& 47 -- 54		& $51.76^{+0.52}_{-0.39}$\\
$\Gamma_c$ 				& 30 -- 300		& $215.4^{+60.3}_{-85.9}$ \\
$\theta_c$ (rad)		& 0.07 -- 0.2	& $0.12^{+0.04}_{-0.03} $ \\
$\theta_v$ (rad)		& 0.1 -- 0.9	& $0.47^{+0.15}_{-0.08} $ \\
$\log (n_0/cm^{-3})$ 	& -5 -- 2		& $-2.68^{+0.88}_{-1.00}$ \\
$\log\epsilon_B$ 		& -5 -- -0.5	& $-4.37^{+1.10}_{-0.48}$ \\
$\log\epsilon_e$ 		& -2 -- -0.5	& $-0.66^{+0.13}_{-0.45}$ \\ 
\hline
\end{tabular}
\end{center}
\tablecomments{We chose a uniform distribution for each prior. The values of the marginalized posterior distributions represent the median the corresponding 16 and 84 percentiles.}
\label{tab:mcmc}
\end{table}

\begin{figure*}
\begin{center}
\includegraphics[width=1\textwidth]{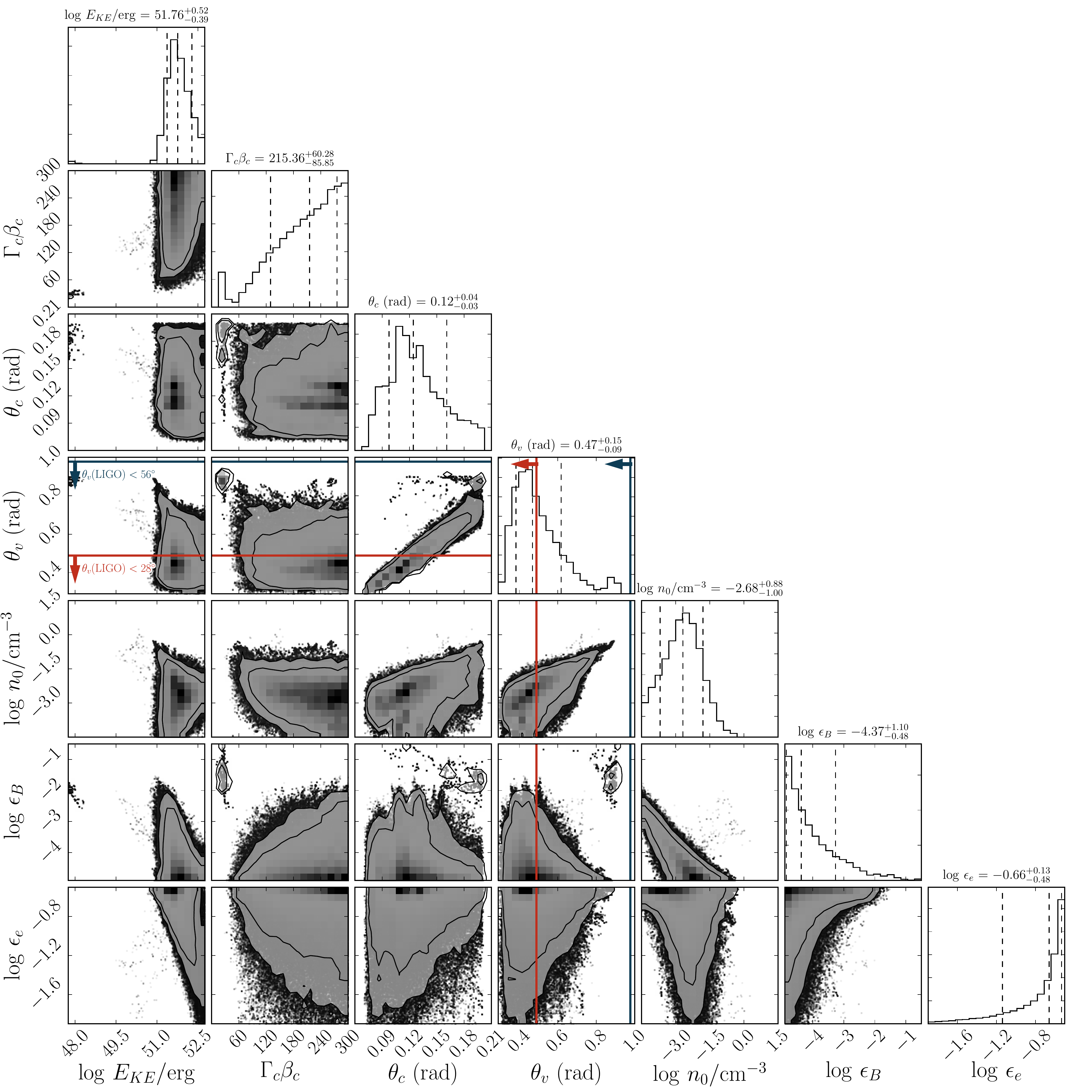}
\end{center}
\caption{Posterior distributions and degeneracies of the model parameters, after removing the burn-in phase. The median values and their $1\sigma$ uncertainties are displayed on top of the marginalized distributions. The blue and red lines display the constraints from the GW signal \citep{LIGO_VIRGO_1}. The tighter constraint of $\theta_v<28^\circ$ was derived using a distance of 42.5 Mpc toward NGC4993, the host of GRB 170817A.}
\label{corner}
\end{figure*}

We used the publicly available affine invariant Markov Chain Monte Carlo parameter estimation code \package{emcee} version 2.2.1 \citep{Foreman2013a} to obtain the bounds of the parameters consistent with the data. We chose a uniform distribution for the prior of each parameter with the ranges displayed in Table \ref{tab:mcmc} and generated $1.5\times 10^6$ realizations of the model ($500$ walkers and $3000$ steps). Our results are presented in Fig. \ref{corner}. To check for convergence, we repeated the simulation multiple times with different initial values of the walkers.

{\commenty{Here, we have only considered observations that were available at the time of the submission of the paper, i.e., up to $158.5$ days. After the submission, later epoch data were released \citep{Dobie:2018zno, DAvanzo:2018zyz}. These show the decline of the radio and X-ray lightcurves. We restrict our analysis to the rising phase of the afterglow in this paper. In a future paper releasing the next set of  GMRT observations, we plan to use the newer data along with a more detailed modeling.}} 

Though the parameter space is degenerate and the data are not highly constraining \citep{Margutti2018a}, we could obtain tight bounds on $E_c, n_0, \theta_c$ and $\theta_v$. $16$ and $84$\% quantiles of the posterior correspond to $2 \times 10^{51} < E_c < 2 \times 10^{52}$. Total energy $E_{\rm tot}$ in the jet 
is given by $\int_0^{2 \pi} d \phi \int_{0}^{\theta_j} d \theta  {\cal E} \sin{\theta}$, which leads to 
\begin{equation}
E_{\rm tot} = \pi E_c \theta^2_c \left\{1 - \exp \left( -\frac{\theta_j^2}{\theta_c^2}\right) \right\}.
\label{etotag}
\end{equation}
Corresponding to the mean of the distribution of $\theta_c$ and $E_c$, $E_{\rm tot} = 2.6 \times 10^{50}$ ergs. The ambient number density, $n_0\sim10^{-2}$--$10^{-4}$\,cm$^{-3}$, is consistent with what is expected for short GRB environments \citep{Fong2015}. Our bounds on $\theta_v$ are consistent with that from the GW analysis from LIGO/Virgo \citep{LIGO_VIRGO_1}. We find that a fairly low fraction ($< 10^{-4}$) of thermal energy density is deposited in the magnetic field, but the fraction going to electrons are toward the maximum possible limit. We notice a tight constraint between $\theta_c $ and $\theta_v$. This comes from the rising slope of the afterglow being tightly constrained by high radio frequencies, as expected from the behavior of the lightcurves seen in Fig.~\ref{variation}. The bulk Lorentz factor is sensitive only to the rise time of afterglow, and hence is not constrained well.  

Our bounds on $\theta_v$ ($27^{+5}_{-5}$ degree) are within  the broad bounds of the LIGO analysis \citep{Abbott:2017xzu}. The combined  LIGO and DES-SHoES bounds presented in \citet{ligosiren2017} are consistent with our posterior. However, the LIGO DES-SHoES bounds along with our posterior will tightly constrain the value of $\theta_v$ between $20^\circ$ and $33^\circ$. The best fit values of $\theta_v$ and $\epsilon_e$ of \citet{Lazzati2017a} are within our posterior bounds, but their $n_0$ is relatively low, where our bounds are at a higher range, and their $\epsilon_B$ is relatively higher than ours. Our posterior is very similar to that of \citep{Troja2018c}. We have slightly tighter bounds on ${\cal E}, \theta_c, \theta_v$, and $n_0$. \citet{Troja2018c} also vary $\theta_j$ and $p$ which we keep fixed. Our bounds on $\epsilon_B$ are broader than theirs and also go down to lower values. The structured-jet parameters \citet{DAvanzo2018a} and \citet{Lyman2018a} have used for the lightcurves they present are well within our posterior bounds.

%

\begin{figure*}
\begin{center}
\includegraphics[width=1\textwidth]{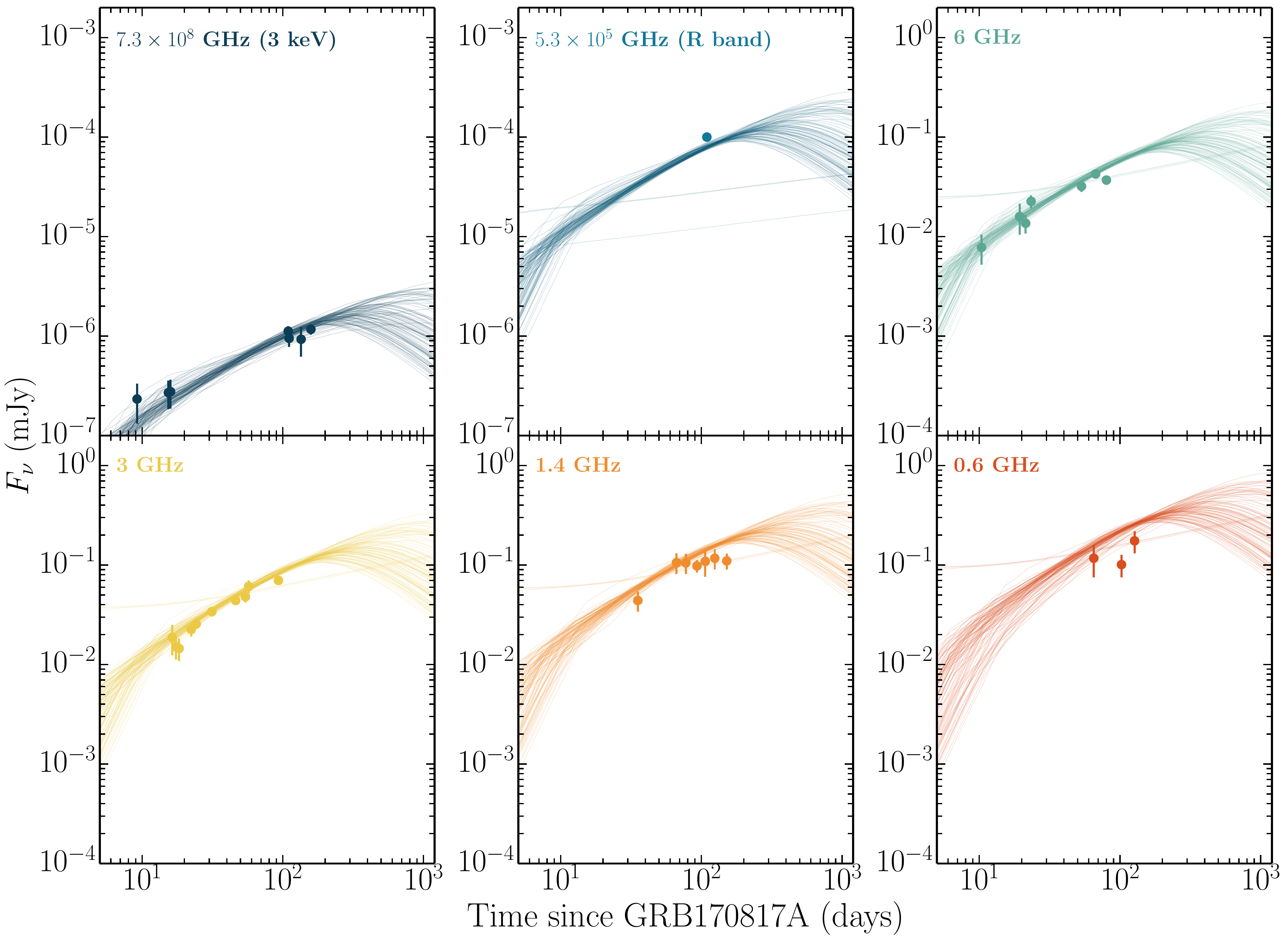} 
\end{center}
\label{finalmodel}
\caption{The $610$ and $1390$~MHz GMRT lightcurves along with higher frequency data from the literature (detections: $\bullet$, upper limits: $\blacktriangledown$). Overlaid are 100 {\commenty{random realizations from the posterior chain from the run where energy fractions are restricted to be $< 0.3$ (Fig \ref{corner}).}} Some lightcurves show small-scale undulations, due to the limited resolution in polar directions. Since the data is dominated by higher radio frequencies, the flatness of the GMRT $1390$~MHz lightcurve is not well reproduced in the model realizations.}
\end{figure*}

In Fig.~\ref{finalmodel}, we present {\sout{100 highest likelihood lightcurves from our model realizations (with similar reduced $\chi^2$ values).}} {\commenty{100 random realizations drawn from the posterior distribution.}} The higher radio frequencies dominate the data, and hence the flatness in GMRT $1390$~MHz is not reproduced well by the models. {\sout{$\theta_v$ in these 100 realizations are broadly divided into two distribution, one around $\sim 21.5^{\circ}$ and the other around $\sim 25^{\circ}$, leading to two classes of lightcurves. The lower $\theta_v$ ones decline earlier, by around $500$~days while the higher $\theta_v$ realization start to decline later.}} Late observations at higher frequencies may agree with the flatness in GMRT L-band lightcurve and refine these predictions.
%
%

{\commenty{We have restricted the upper bounds of the prior distributions of the micro-physical parameters $\epsilon_e$ and $\epsilon_B$ to $0.3$ (Table \ref{tab:mcmc}), allowing at most a third of the shock generated thermal energy in the magnetic field and in the non-thermal electrons each. However, since the posterior of $\epsilon_e$ shows a preference toward higher values, we ran a test simulation by extending the prior distribution to unphysical values ($\epsilon_{e,B} > 1$). The resultant posterior of $\epsilon_e$ peaks at a relatively high value of $0.9$, and the $84$ percentile quantile extends to an unphysical value of $1.9$. In addition, extending the priors of $\epsilon_e$ and $\epsilon_B$ demanded extending the parameter spaces of $\theta_v$ and $n_0$ as well. The resultant posterior of $\theta_v$ is very different from the case where $\epsilon_{e,B} < 0.3$, and  is not in agreement with the Ligo/Virgo limit of $\theta_v < 28^{\circ}$. Posteriors of two other parameters, $n_0$ and $\epsilon_B$ also change remarkably in this case, to clear multi-modal distributions. The posterior of $\epsilon_B$ peaks at a  fairly low $10^{-7}$. We present the final posterior distribution in Fig \ref{secondcorner} and Tab \ref{tab:secondpost}. For further analysis we do not consider this run, as it favors unphysical ranges of microphysical parameters and values of $\theta_v$ above the Ligo/Virgo limit. However, this behaviour of the posterior may be a hint of additional elements required in the afterglow model. On the other hand, it also may be an indication that the amount of data is still too small to accurately constrain the model parameters. A more detailed analysis will be presented in a forthcoming paper releasing our next set of data.}
\begin{figure*}
\begin{center}
\includegraphics[width=1\textwidth]{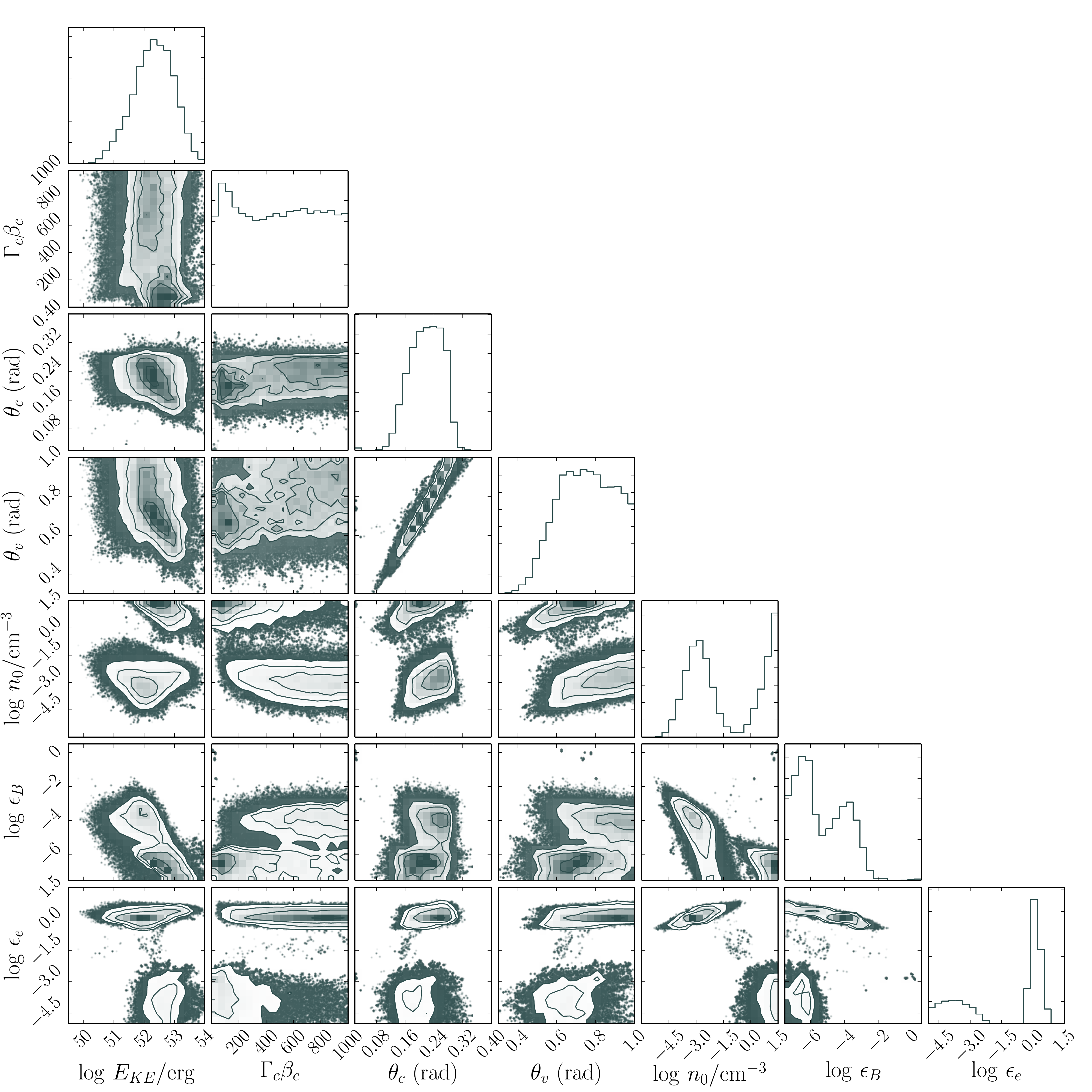}
\end{center}
\caption{{\commenty{Posterior distribution for the MCMC run where the prior ranges of the microphysical parameters were not restricted to be below unity. Posterior of $\theta_v$ and $n_0$ have also changed considerably from Fig \ref{corner}. Bimodal distributions appear in $n_0, \epsilon_B,$ and $\epsilon_e$. Likelihood peaks at unphysical or unrealistic values for microphysical parameters. Additional elements in the model, or more data, or both may require to constrain the posterior better. Such an analysis is beyond the scope of this paper and we postpone it for future.}}}
\label{secondcorner}
\end{figure*}
\begin{table}
\caption{Posterior distributions for unrestricted $\epsilon_e$ prior}
\begin{center}
\begin{tabular}{cc}
Parameter 					& Posterior\\ 
\hline
$\log (E_c/{\rm erg})$ 				& $52.33^{+0.64}_{-0.70}$\\
$\Gamma_c$ 					& $493.6^{+343.4}_{-353.5}$ \\
$\theta_c$ (rad)			& $0.22^{+0.05}_{-0.05} $ \\
$\theta_v$ (rad)			& $0.75^{+0.16}_{-0.15} $ \\
$\log (n_0/cm^{-3})$ 			& $-2.27^{+3.40}_{-1.11}$ \\
$\log\epsilon_B$ 			& $-5.77^{+2.08}_{-1.07}$ \\
$\log\epsilon_e$ 		& $-0.09^{+0.37}_{-3.86}$ \\ 
\hline
\end{tabular}
\end{center}
\tablecomments{We chose a uniform distribution for each prior. The values of the marginalized posterior distributions represent the median the corresponding 16 and 84 percentiles.}
\label{tab:secondpost}
\end{table}

%
%
%
%
%

\section{Constraints from prompt emission}
In the off-axis structured-jet scenario, the properties of the observed prompt emission, particularly the isotropic equivalent energy ($E_{\rm iso}^{\gamma}$), is sensitive to the jet bulk Lorentz factor ($\Gamma_c$), jet structure parameter ($\theta_c$), and viewing angle ($\theta_v$) \citep{donaghy05,Yamazaki03}. Assuming that the burst and the afterglow are produced by an off-axis relativistic jet, further constraints on $\Gamma_c$, $\theta_c$, and $\theta_v$ can be arrived at in conjunction with the afterglow parameter space. We have carried out the same analysis in paper-I for uniform top-hat jet. Here we extend it to the Gaussian structured jet. 

The observed GRB flux, from which the $E_{\rm iso}^{\gamma}$ is derived, is the intensity integrated over the surface of the jet visible to the observer. The flux depends on the viewing angle $\theta_v$, the energy content of the jet, and the bulk Lorentz factor. In order to obtain $E_{\rm iso}^{\gamma}$, we follow the framework developed by \citet{Salafia:2015vla}. They derive that the isotropic equivalent energy in prompt emission observed by an off-axis observer is,
\begin{equation}
E_{\rm iso}^{\gamma} (\theta_v) = \int d\Omega \frac{\delta^3(\alpha)}{\Gamma(\theta)} u^{\gamma}(\theta) ,
\label{eqetot}
\end{equation}
where, $\delta (\alpha)$ is the Doppler factor given by $1/\Gamma (1-\beta \cos{\alpha})$ and $u^{\gamma}(\theta)$ is the energy per solid angle in prompt emission. The angle $\alpha$ entering in the expression of $\delta$ is the same as defined in $\S$\ref{sec:model}; the angle between the observer line of sight and the normal to jet surface at $\theta, \phi$. 

We consider, $u^{\gamma}(\theta)$ to follow the same functional dependence as ${\cal E}$, energy per solid angle in the afterglow blast wave (Eq. \ref{eq1}). 
\begin{equation}
	u^{\gamma}(\theta)= u_c\exp{-\frac{\theta^2}{\theta_c^2}}.
\end{equation}
In order to obtain the normalization $u_c$, which gives the energy per solid angle at the jet axis ($\theta=0$), we  assume that the kinetic energy budget in the afterglow, $E_{\rm tot}$, given in Eq. \ref{etotag}, and the total energy radiated away in prompt emission as measured by an on-axis observer are related by a factor $\zeta$. This assumption is motivated by a constant $\gamma$-ray efficiency \citep{cenko2011} seen in long GRBs detected by $\gamma$-ray triggers, where the observer is very likely to be aligned close to the axis of the jet. 

Total isotropic equivalent energy in prompt emission as measured by an on-axis observer is $E_{\rm iso}^{\gamma} (\theta_v=0)$. Therefore, to obtain $u_c$,
\begin{equation}
u_c \int_0^{\theta_j} d \Omega \exp\left(-\frac{\theta^2}{\theta_c^2}\right) \frac{\delta^3(\theta_v =0)}{\Gamma (\theta)}  = \zeta \frac{E_{\rm tot}}{(1-\cos{\theta_j})}.
\label{equc}
\end{equation}
The factor $(1-\cos{\theta_j})$ converts the afterglow energy $E_{\rm tot}$ to isotropic equivalent energy. 

We consider {\commenty{5000 random realizations of the the posterior chain}} and see whether a given realization can reproduce the observed $E_{\rm iso}^{\gamma}$ \citep{Goldstein2017b} with reasonable values of $\zeta$. From \citet{cenko2011}, a variation in $\zeta$ from $0.05$ to $40$ is commonly seen, with a mean value being $\sim 4$.

For the given realization, we first obtain $u_c$ using Eq. \ref{equc} after replacing $\theta_v$ by $0$ and then proceed to obtain $E_{\rm iso}^{\gamma} (\theta_v)$ using Eq. \ref{eqetot}. In Figure \ref{figprompt}, we display our results for $\zeta = 1$ and $\zeta = 0.1$. We find that the observed Fermi $E_{\rm iso}^{\gamma}$ can be reproduced for a good fraction of the posterior if the energy budget in prompt emission is of an order of magnitude lower than that in the afterglow. For $\zeta = 1$, i.e., similar energies in the afterglow and prompt emission (i.e., for $50$ \% $\gamma$-ray efficiency), only a few low bulk Lorentz factor solutions can reproduce the observed $E_{\rm iso}^{\gamma}$. However, a $\zeta \sim 0.1$ is not unusual of ordinary GRBs \citep{cenko2011}, and is also supportive of a low efficiency process like internal shocks. {\commenty{This analysis shows that a relativistic structured outflow is successful in describing both the prompt and the afterglow observations of GRB 170817A, albeit the posterior distribution of $\epsilon_e$ currently prefers extremely high values.}}

\begin{figure*}
\begin{center}
\includegraphics[width=1\columnwidth]{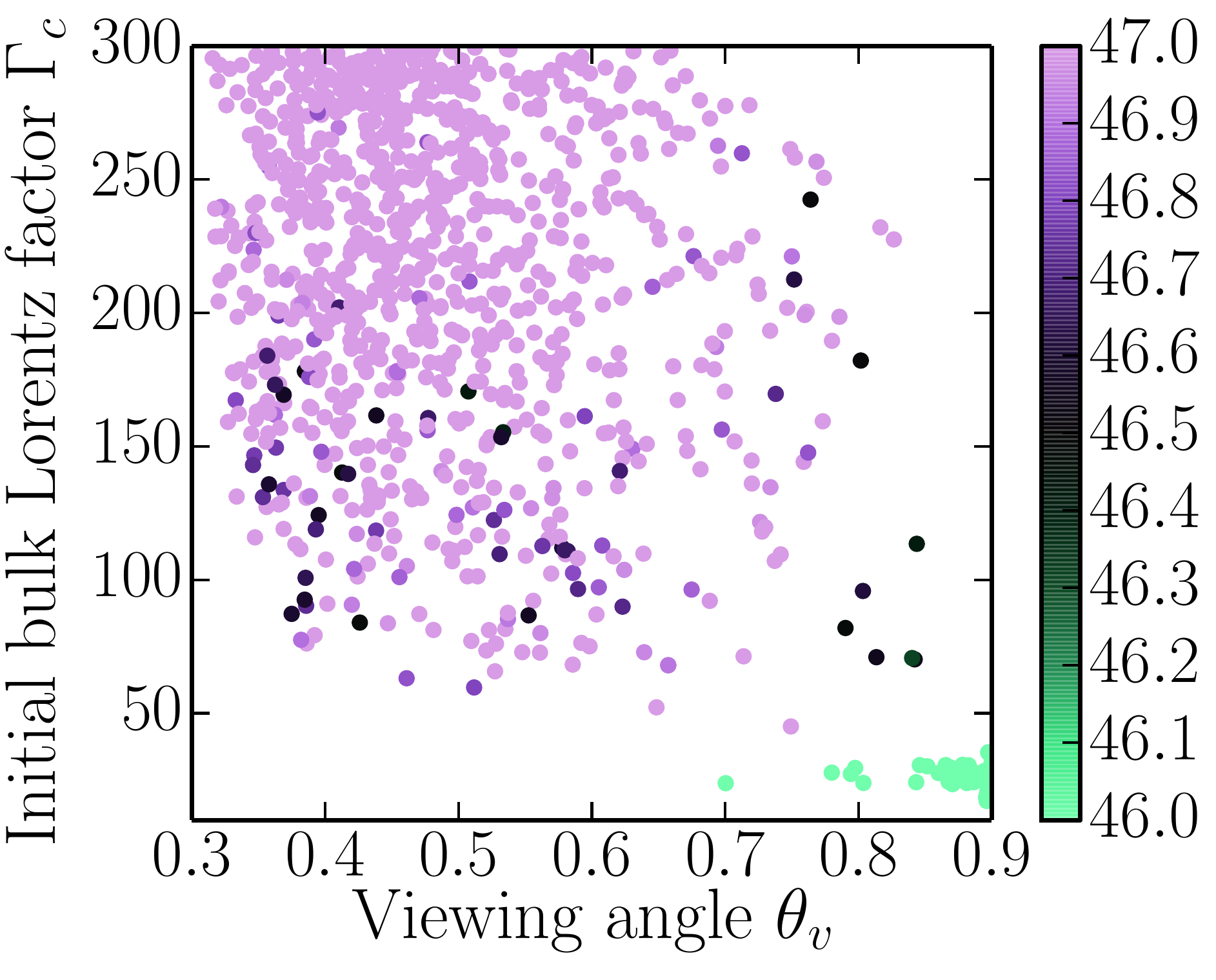} 
\includegraphics[width=1\columnwidth]{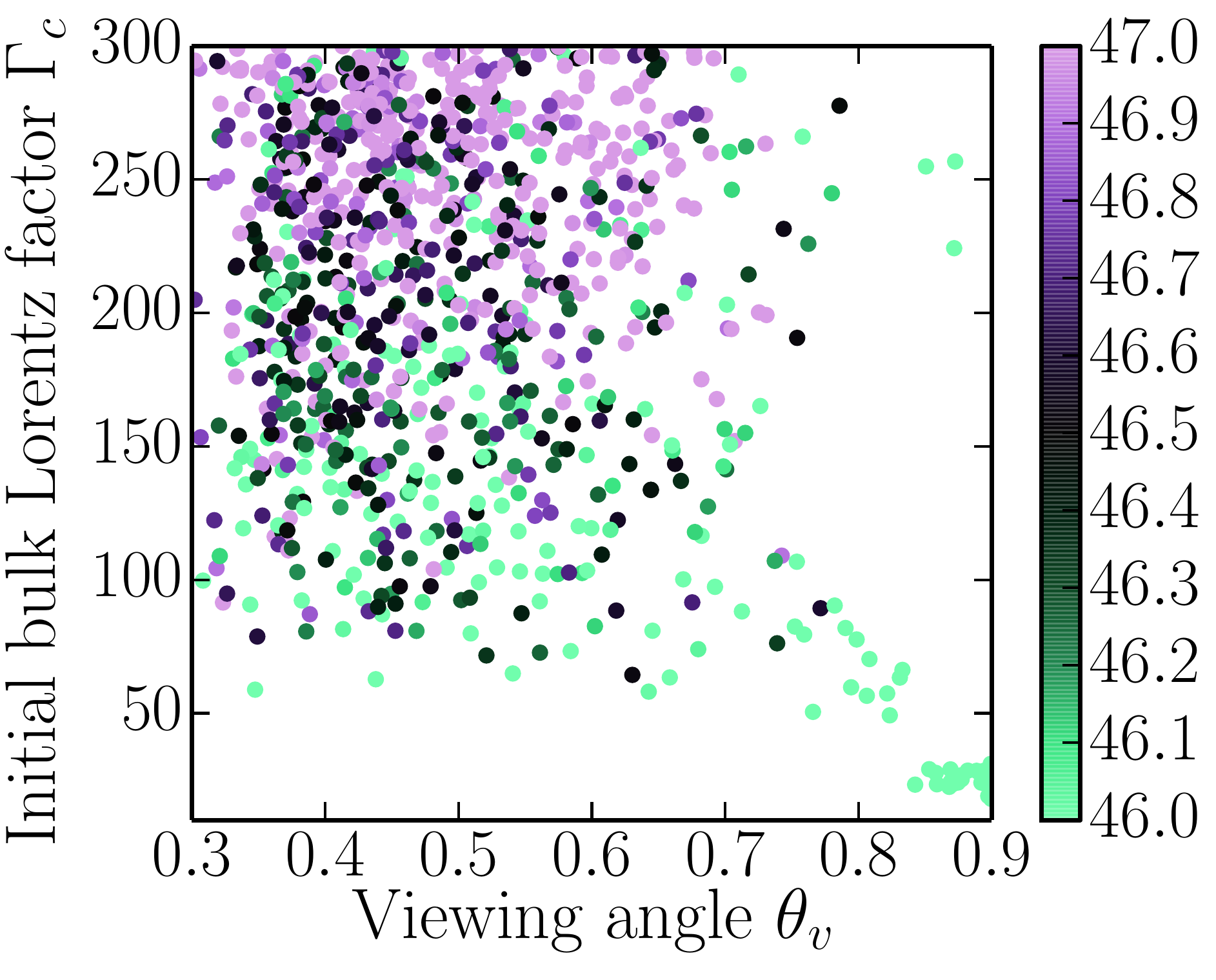} 
\caption{$E_{\rm iso}^{\gamma}$ reproduced by {\commenty{5000 random realizations of the afterglow solutions}}. The points are color-coded with the numerically estimated value of $E_{\rm iso}^{\gamma}$. While the afterglow solutions have 4 parameters relevant to the prompt emission ($E_c, \theta_c, \theta_v, \Gamma_c$), we have only shown two dimensions ($\Gamma_c$ and $\theta_v$) here. {\commenty{The anti-correlation, i.e., a higher $\Gamma_c$ requires a lower $\theta_v$ to be able to reproduce a given $E_{\rm iso}^\gamma$ and vice versa, is obvious in the figure.}}  The Fermi observed $E_{\rm iso}^{\gamma} = (3.08 \pm 0.72) \times 10^{46}$~erg \citep{Goldstein2017b} (between $46.4$ and $46.6$ in log-scale).  First panel is for $\zeta=1$ and the second panel is for $\zeta=0.1$.  A highly relativistic structured outflow can well describe both the prompt and afterglow observations. Nevertheless, the prompt emission seems to have carried relatively less amount of energy than the afterglow.}
\label{figprompt}
\end{center}
\end{figure*}

\section{Summary}\label{sec:summary}
We present the low radio frequency observations of the afterglow of GRB 170817A/GW170817 with the Giant Metrewave Radio telescope. We began detecting the afterglow at $1390$~MHz starting from $\sim 65$~days since burst and is continuing to follow it up in low radio frequencies. We present $1390$~MHz observations up to $152$~days since burst. The lightcurve is particularly flat, which may indicate a slow turn-over in the flux evolution \footnote{\commenty{This claim is supported by later observations published while the paper was in revision \citep{Dobie:2018zno}.}} 

We interpreted the multi-wavelength afterglow in the framework of a structured jet with a Gaussian velocity and energy profile. Bounds on the jet energy, angular structure, observer's viewing angle, and ambient density are  obtained through an MCMC parameter estimation. The energy per solid angle at the jet axis is $5.8^{+13.3}_{-3.4} \times 10^{51}$~erg, $\theta_c$ is $6.9^{+2.3}_{-1.5}$ degree, $\theta_v$ is $27^{+8}_{-5}$ degree, and ambient density is $0.002^{+0.014}_{-0.002}~{\rm cm}^{-3}$. While the initial
bulk Lorentz factor could not be well constrained, a relativistic flow
with $\Gamma$ of a few hundred close to the jet axis is perfectly acceptable. These parameters are consistent with that of typical short-duration GRBs. We find that isotropic energy observed in the prompt emission can be reproduced if the total energy budget in the prompt emission is an order of magnitude lower than that in the afterglow. Such a difference in energy content is not unusual for GRBs. Our analysis supports the view that GRB 170817A is similar to standard GRBs, with typical energetics and bulk Lorentz factors, {\commenty{except that it is viewed at an angle less probable for $\gamma$-ray triggered events. Seen at such extreme angles, presence of an angular structure in the outflow has become evident in case of GRB 170817A.}} 
\section*{Acknowledgments}
We thank the staff of the GMRT that made these observations possible. GMRT is run by the National Centre for Radio Astrophysics of the Tata Institute of Fundamental Research.

We thank Avishay Gal-Yam and Doron Kushnir for using the high-performance computing (HPC) facility WEXAC at the Weizmann Institute of Science. The Weizmann HPC facility is partly supported by the Israel Atomic Energy Commission - The Council for Higher Education - Pazi Foundation and partly by a research grant from The Abramson Family Center for Young Scientists.

Development of the \package{BOXFIT} code was supported in part by NASA through grant NNX10AF62G issued through the Astrophysics Theory Program and by the NSF through grant AST-1009863. Simulations for \package{BOXFIT} version 2 have been carried out in part on the computing facilities of the Computational Center for Particle and Astrophysics (C2PAP) of the research cooperation "Excellence Cluster Universe" in Garching, Germany. This work made use of the IAA-CSIC high performance (HPC) and throughput (HTC) computing infrastructure.

RL acknowledges support from the grant EMR/2016/007127 from Dept. of Science and Technology, India. RL thanks M Govindankutty (IIST, Trivandrum) for generously providing computing facilities and ICTS, Bangalore for hospitality and computing facilities. JB acknowledges support from the CONICYT-Chile grant Basal-CATAPFB-06/2007 and FONDECYT Postdoctorados 3160439. RS-R acknwoledges support from ASI (Italian Space Agency) through the Contract n. 2015-046-R.0 and from European Union Horizon 2020 Programme under the AHEAD project (grant agreement n. 654215). SS acknowledges support from the Feinberg Graduate School. 

RS-R thanks for the excellent support and advice from Rafael Parra (Computing Center, IAA-CSIC).

\bibliography{mybib}

\end{document}